\documentclass[pdftex,twocolumn,epjc3]{svjour3} 
\smartqed  
\pdfoutput=1 

\RequirePackage{graphicx}
\RequirePackage{mathptmx}      
\RequirePackage[numbers,sort&compress]{natbib}
\RequirePackage[colorlinks,citecolor=blue,urlcolor=blue,linkcolor=blue]{hyperref}

\usepackage[usenames,dvipsnames]{xcolor}
\usepackage[normalem]{ulem}     
\usepackage{upgreek}
\usepackage[version=4]{mhchem}  
\usepackage{xspace}             
\usepackage{fancyhdr}
\usepackage[switch]{lineno}     

\newcommand{\ton}{t}
\newcommand{\ntpcdiameter}{$96~\mathrm{cm}$\xspace}
\newcommand{\ntpclength}{$97~\mathrm{cm}$\xspace}
\newcommand{\vzbb}{$0\nu\beta\beta$\xspace}
\newcommand{\vtbb}{$2\nu\beta\beta$\xspace}

\newcommand{\bologna}{Department of Physics and Astronomy, University of Bologna and INFN-Bologna, 40126 Bologna, Italy}

\newcommand{\chicago}{Department of Physics \& Kavli Institute for Cosmological Physics, University of Chicago, Chicago, IL 60637, USA}

\newcommand{\coimbra}{LIBPhys, Department of Physics, University of Coimbra, 3004-516 Coimbra, Portugal}

\newcommand{\columbia}{Physics Department, Columbia University, New York, NY 10027, USA}

\newcommand{\lngs}{INFN-Laboratori Nazionali del Gran Sasso and Gran Sasso Science Institute, 67100 L'Aquila, Italy}

\newcommand{\mainz}{Institut f\"ur Physik \& Exzellenzcluster PRISMA, Johannes Gutenberg-Universit\"at Mainz, 55099 Mainz, Germany}

\newcommand{\heidelberg}{Max-Planck-Institut f\"ur Kernphysik, 69117 Heidelberg, Germany}

\newcommand{\munster}{Institut f\"ur Kernphysik, Westf\"alische Wilhelms-Universit\"at M\"unster, 48149 M\"unster, Germany}

\newcommand{\nikhef}{Nikhef and the University of Amsterdam, Science Park, 1098XG Amsterdam, Netherlands}

\newcommand{\nyuad}{New York University Abu Dhabi, Abu Dhabi, United Arab Emirates}

\newcommand{\purdue}{Department of Physics and Astronomy, Purdue University, West Lafayette, IN 47907, USA}

\newcommand{\rpi}{Department of Physics, Applied Physics and Astronomy, Rensselaer Polytechnic Institute, Troy, NY 12180, USA}

\newcommand{\rice}{Department of Physics and Astronomy, Rice University, Houston, TX 77005, USA}

\newcommand{\stockholm}{Oskar Klein Centre, Department of Physics, Stockholm University, AlbaNova, Stockholm SE-10691, Sweden}

\newcommand{\subatech}{SUBATECH, IMT Atlantique, CNRS/IN2P3, Universit\'e de Nantes, Nantes 44307, France}

\newcommand{\torino}{INAF-Astrophysical Observatory of Torino, Department of Physics, University  of  Torino  and  INFN-Torino,  10125  Torino,  Italy}

\newcommand{\ucla}{Physics \& Astronomy Department, University of California, Los Angeles, CA 90095, USA}

\newcommand{\ucsd}{Department of Physics, University of California San Diego, La Jolla, CA 92093, USA}

\newcommand{\wis}{Department of Particle Physics and Astrophysics, Weizmann Institute of Science, Rehovot 7610001, Israel}

\newcommand{\zurich}{Physik-Institut, University of Zurich, 8057  Zurich, Switzerland}

\newcommand{\paris}{LPNHE, Sorbonne Universit\'{e}, Universit\'{e} de Paris, CNRS/IN2P3, Paris, France}

\newcommand{\freiburg}{Physikalisches Institut, Universit\"at Freiburg, 79104 Freiburg, Germany}

\newcommand{\lal}{Universit\'{e} Paris-Saclay, CNRS/IN2P3, IJCLab, 91405 Orsay, France}

\newcommand{\naples}{Department of Physics ``Ettore Pancini'', University of Napoli and INFN-Napoli, 80126 Napoli, Italy} 

\newcommand{\nagoya}{Kobayashi-Maskawa Institute for the Origin of Particles and the Universe, Nagoya University, Furo-cho, Chikusa-ku, Nagoya, Aichi 464-8602, Japan}

\newcommand{\laquila}{Department of Physics and Chemistry, University of L'Aquila, 67100 L'Aquila, Italy}

\newcommand{\tokyo}{Kamioka Observatory, Institute for Cosmic Ray Research, and Kavli Institute for the Physics and Mathematics of the Universe (WPI), the University of Tokyo, Higashi-Mozumi, Kamioka, Hida, Gifu 506-1205, Japan}

\newcommand{\ferrara}{INFN, Sez. di Ferrara and Dip. di Fisica e Scienze della Terra, Università di Ferrara, via G. Saragat 1, Edificio C, I-44122 Ferrara (FE), Italy}

\newcommand{\suny}{Simons Center for Geometry and Physics and C. N. Yang Institute for Theoretical Physics, SUNY, Stony Brook, NY, USA}

\newcommand{\utrecht}{Institute for Subatomic Physics, Utrecht University, Utrecht, Netherlands}

\newcommand{\spacenagoya}{Institute for Space-Earth Environmental Research, Nagoya University, Nagoya, Aichi 464-8601, Japan}

\newcommand{\coimbrapoli}{Coimbra Polytechnic - ISEC, Coimbra, Portugal}

\newcommand{\kobe}{Department of Physics, Kobe University, Kobe, Hyogo 657-8501, Japan}

\journalname{Eur. Phys. J. C}
\title{Energy resolution and linearity of XENON1T in the MeV energy range}

\author{E.~Aprile\thanksref{addr1}
        \and
        J.~Aalbers\thanksref{addr2}
        \and
        F.~Agostini\thanksref{addr3}
        \and
        M.~Alfonsi\thanksref{addr4}
        \and
        L.~Althueser\thanksref{addr5}
        \and
        F.~D.~Amaro\thanksref{addr6} 
        \and
        V.~C.~Antochi\thanksref{addr2}
        \and
        E.~Angelino\thanksref{addr7}
        \and
        J.~Angevaare\thanksref{addr11} 
        \and
        F.~Arneodo\thanksref{addr8}
% ...
        (XENON Collaboration\thanksref{t1}).
}
\authorrunning{XENON Collaboration}
\thankstext{also1}{Also at \suny}
\thankstext{also2}{Also at \utrecht}
\thankstext{also3}{Also at \spacenagoya}
\thankstext{also4}{Also at \coimbrapoli}
\thankstext{also5}{At \ferrara}
\thankstext{e1}{\texttt{chiara.capelli@physik.uzh.ch}}
\thankstext{e2}{\texttt{tianyu.zhu@columbia.edu}}
\thankstext{t1}{\texttt{xenon@lngs.infn.it}}
\institute{\columbia\label{addr1} 
        \and
        \stockholm\label{addr2} 
        \and
        \bologna\label{addr3} 
        \and
        \mainz\label{addr4} 
        \and
        \munster\label{addr5} 
        \and
        \coimbra\label{addr6} 
        \and
        \torino\label{addr7} 
        \and
        \nikhef\label{addr11} 
        \and
        \nyuad\label{addr8} 
% ...
}

\begin{document}
\date{}
\maketitle

\begin{abstract}
Xenon dual-phase time projection chambers designed to search for Weakly Interacting Massive Particles have so far shown a relative energy resolution which degrades with energy above $\sim$200 keV due to the saturation effects. This has limited their sensitivity in the search for rare events like the neutrinoless double-beta decay of $^{136}$Xe at its $Q$-value, $Q_{\beta\beta}\simeq$ 2.46\,MeV. For the XENON1T dual-phase time projection chamber, we demonstrate that the relative energy resolution at 1\,$\sigma/\mu$ is as low as (0.80$\pm$0.02)\,\% in its one-ton fiducial mass, and for single-site interactions at $Q_{\beta\beta}$. We also present a new signal correction method to rectify the saturation effects of the signal readout system, resulting in more accurate position reconstruction and indirectly improving the energy resolution. The very good result achieved in XENON1T opens up new windows for the xenon dual-phase dark matter detectors to simultaneously search for other rare events.


\end{abstract}

\flushbottom

\section{Introduction}
\label{sec:intro}

The search for dark matter and the investigation of the fundamental nature of neutrinos are two outstanding endeavours in contemporary physics. The dual-phase xenon time projection chambers (TPCs), led by the XENON1T experiment, has achieved to date the most stringent upper limits on spin-independent~\cite{Aprile:2018dbl} and spin-dependent neutron~\cite{Aprile:2019dbj} interactions for WIMPs with mass above 6\,GeV/c$^2$, as well as for sub-GeV dark matter particles~\cite{Aprile:2019xxb}.
XENON1T uses xenon containing $^{136}$Xe with isotopic abundance of 8.49\%, it can therefore also search for the neutrinoless double-beta decay (\vzbb) at its $Q$-value,  $Q_{\beta\beta}$ = (2457.83$\pm$0.37)\,keV \cite{PhysRevLett.98.053003}.
A detection of \vzbb would establish the Majorana nature of neutrinos and demonstrate lepton number violation by two units. The experimental signature of \vzbb is a mono-energetic peak at $Q_{\beta\beta}$, at the falling end of the continuous energy spectrum of the two-neutrino double beta decay (\vtbb) standard model process. The \vzbb half-life sensitivity depends on the total detection efficiency, $\epsilon$, the isotopic abundance, $n_{136}$, the atomic mass number $m_A$ of $^{136}$Xe, and the total exposure $M\cdot t$, where $M$ is the fiducial mass, and $t$ is the livetime of the measurement.
In the absence of signal events, in an energy interval $\Delta E$ around $Q_{\beta\beta}$, the 90\% C.L. limit on the half-life can be expressed as 
\begin{linenomath}
\begin{equation}
T_{1/2}^{0\nu} > \frac{\ln 2}{1.64} \frac{N_A}{m_A} \cdot \epsilon \cdot n_{136} \cdot \frac{M \cdot t}{\sqrt{n_B}} \propto \sqrt{\frac{M \cdot t}{B \cdot \Delta E}}\text{ ,}
\end{equation}
\end{linenomath} where $N_A$ is Avogadro's number, $n_B$ is the number of expected background events and $B$ is the background rate in the energy interval~\cite{DellOro:2016tmg}. A good energy resolution is fundamental to minimize the region $\Delta E$, thus enhancing the experimental sensitivity. This paper describes several improvements to the signal reconstruction algorithms for XENON1T, leading to excellent energy linearity and resolution at $Q_{\beta\beta}$.

\section{The XENON1T experiment}
\label{sec:xenon1t}

The XENON1T detector is a dual-phase xenon TPC which consists of a \ntpclength length and \ntpcdiameter diameter cylindrical active detection volume containing 2\,t of ultra-pure liquid xenon (LXe) out of a total of 3.2\,t in the detector. Two arrays of Hamamatsu R11410-21 3" photomultiplier tubes (PMTs)~\cite{Aprile:2015lha} are arranged above and below the sensitive volume of the TPC. The side walls of the cylindrical volume are PTFE reflectors that enhance the light collection efficiency. 
Energy depositions from interactions in the LXe produce both scintillation photons and ionization electrons. The scintillation light signal (S1) is promptly detected by the PMTs. A grounded electrode, the gate, placed just $\sim$2.5\,mm below the liquid-gas interface, and a cathode placed at the bottom of the TPC produce an average electric field of 81\,V/cm to drift electrons produced in the liquid upwards with a drift velocity of 1.33\,mm/$\mu$s. An anode is placed 5\,mm above the gate and the 8.1\,kV/cm electric field between them extracts electrons into the gaseous xenon with an electron extraction efficiency calculated to be 96\,\%~\cite{Aprile:2017aty}. Here the electrons produce proportional scintillation light signal (S2), which is also detected by the PMTs. The time delay between S1 and S2 is used to reconstruct the interaction depth ($z$ position) with a resolution down to 0.5\,mm. The distribution of the S2 light on the top PMT array is used to reconstruct the $x$-$y$ position, reaching a resolution of 8\,mm for S2 values above $10^{3}$ photo-electrons (PE)~\cite{Aprile:2019bbb}. The PMTs have an average quantum efficiency of 34.5\% and channel-dependent gains of (1.0-5.0)$\times10^6$\,\cite{Barrow:2016doe}. The signals are guided to Phillips 776 amplifiers that provide an additional amplification factor of 10. The output of the amplifiers is sent to CAEN V1724 waveform digitizer modules to record the signals at a sampling rate of 100\,MHz with a 2.25\,V dynamic range, a 40\,MHz input bandwidth, and 14\,-bit resolution. The data acquisition system is described in detail in \cite{Aprile:2019cee}.
 
\section{Signal reconstruction techniques}
\label{sec:reconstruction}
The data processing in XENON1T is performed with the modular software package Processor for Analyzing XENON (PAX)\cite{Aprile:2019bbb}\cite{xenon_collaboration_2018_1195785}. This section describes several improvements to the low-level signal reconstruction routines of PAX for the dark matter search in order to optimize detector performance up to the MeV energy range.

\subsection{Waveform saturation correction}
\label{subsec:saturation_correction}
\begin{figure}[h]
    \centering
    \includegraphics[width=0.46\textwidth]{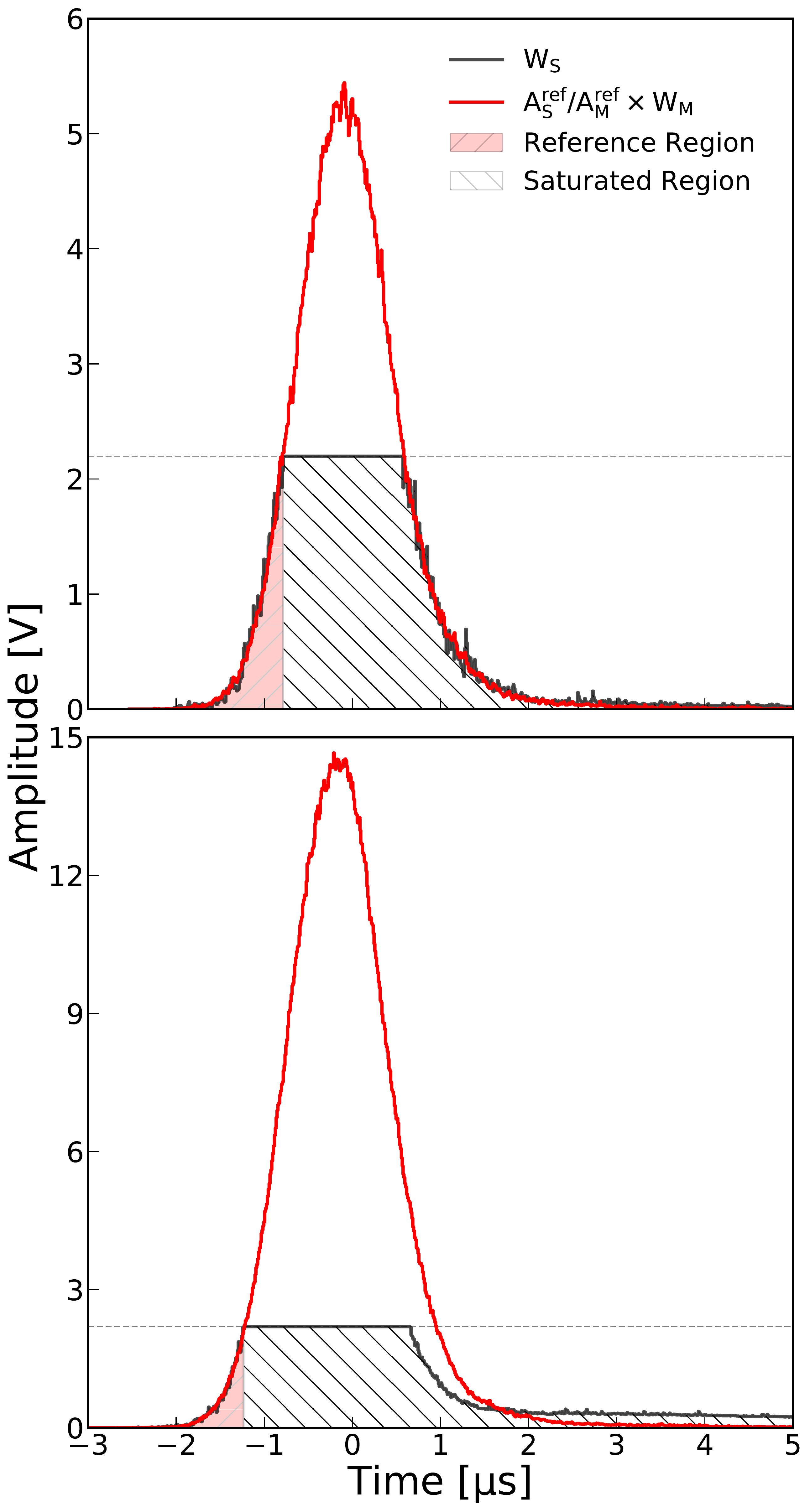}
    \caption{Examples of saturated WFs from two S2s with a size of about $2\times10^{5}$\,PE (top) and $10^{6}$\,PE (bottom). Each panel shows a WF (black) in one channel centred to time zero. Both WFs are truncated due to the range of the digitizer. The WF model, obtained from the sum of non-saturated WFs, is scaled and overlaid in the plot (red). The red shaded region each covers 1\,$\mu$s before the first truncated sample and used as a reference region, while the hatched region from the first truncated sample to the end of the pulse covers the range where WFs are corrected as the scaled WF model.}
    \label{fig:pulse_desaturation}
\end{figure}

XENON1T, designed for dark matter searches, features a signal readout system optimized to amplify and detect tiny signals down to single PE from individual PMTs\,\cite{Aprile:2019cee}. For interactions with energies $\sim$1 MeV, several components, including the PMT voltage divider circuits, the amplifiers and the digitizers will saturate, resulting in distorted output S2 signals. A correction for saturation effects is thus critical for reconstructing signals at MeV energies with sufficient energy resolution for \vzbb searches. The digitizer saturation occurs at energies above $\sim200$\,keV, corresponding to a total S2 signal on the order of $10^{5}$\,PE and as large as $\sim5\times10^{4}$\,PE in the PMT with the largest signal. The exact energy saturating the digitizers varies according to the location of the interaction. Such signals exceed the 2.25\,V dynamic range of the digitizers and result in truncated waveforms (WFs). Non-linear responses of the PMT voltage divider circuits and the amplifiers are expected to occur at a higher energy of $\sim1$\,MeV, corresponding to an S2 signal on the order of $10^{6}$\,PE. For these events, the analog (or the pre-digitizer) signals are distorted and no longer proportional to the number of initial photons~\cite{JWulf_thesis}. Examples of S2 signals corresponding to those two cases are shown in Fig.\,\ref{fig:pulse_desaturation}.

The correction method described in this section is based on the temporal and spatial characteristics of the S2. The S2 has a wide (at least 0.5\,$\mu$s at 1$\sigma$ width) and nearly identical temporal distribution across all channels because the proportional scintillation light is produced for the duration of electrons drifting from the liquid-gas interface to the anode. 
Additionally, this light is produced $\simeq7$cm\ below the top PMT array, with the majority of it hitting a few PMTs. Some PMTs, especially those on the top array and away from the $x$-$y$ coordinate of the S2, remain unsaturated. The pulse shape of S2 signals in those non-saturated channels are used to correct signals in the saturated channels. The correction procedure applies to individual peaks and is as follows:
\begin{enumerate}\label{saturation_correction}
    \item Sorting all S2 WFs into two classes: saturated and non-saturated, based on whether the WF reaches the limit of the dynamic range of the digitizers. 
    \item All non-saturated WFs are summed together to get a WF model denoted as $\rm{W}_{\rm{M}}$. This WF model is an unbiased estimate of the S2 WF shape.
    \item For each saturated WF denoted as $\rm{W}_{\rm{S}}$, the region before the first saturated sample is used as a reference region. We denote the integral of $\rm{W}_{\rm{S}}$ and of $\rm{W}_{\rm{M}}$ over the reference region as $\rm{A}^{\rm{ref}}_{\rm{S}}$ and $\rm{A}^{\rm{ref}}_{\rm{M}}$, respectively.
    \item Each saturated WF is corrected as $\rm{A}^{\rm{ref}}_{\rm{S}}/\rm{A}^{\rm{ref}}_{\rm{M}}\times\rm{W}_{\rm{M}}$ after the reference region.
    \item The correction is applied to the region from the first saturated sample to the last sample of the WFs, 1\,$\mu$s after the WFs fall below the channel-specific trigger thresholds. In rare cases when another peak starts before the end of the WFs, the corrections stop right before the second peak.
\end{enumerate}
 
Two representative examples of S2 each with a WF in a saturated channel are shown with the scaled model $\rm{W}_{\rm{M}}$ overlaid in Fig.\,\ref{fig:pulse_desaturation}. For the S2 $\sim 2\times10^{5}$\,PE shown in the top panel, the analog signal is not distorted and the falling edge of the $\rm{W}_{\rm{S}}$ agrees well with $\rm{W}_{\rm{M}}$. This is not the case with a larger S2 $\sim 10^{6}$\,PE, as shown in the bottom panel. Here, $\rm{W}_{\rm{M}}$ does not match the falling edge of the $\rm{W}_{\rm{S}}$. In particular, the undershoot of $\rm{W}_{\rm{S}}$ is not caused by the saturation of the digitizers but by saturation of the PMTs or amplifiers. However, the $\rm{W}_{\rm{M}}$ is still a valid model because the signal in the reference region is not yet large enough to induce saturation in those two components. The overshoot present on the right side is instead mostly due to secondary signals, as it will be clarified in Sec.~\ref{subsec:peak_indentification}. In order to rectify all saturation effects, the correction is extended to the last sample of $\rm{W}_{\rm{S}}$ in all cases.
In addition to the impact on the energy reconstruction, the saturation correction also notably affects the position reconstruction and thus the spatial correction for the S1 and S2 signals, as shown in Sec.~\ref{pos_rec}.

Unlike the S2, the S1 light is more evenly distributed among all PMTs and it is not amplified in the gas region. As a result, S1 signals from electronic recoils have negligible saturation, even for events with energies in MeV region. In addition to this, the scintillation photons are produced on much shorter timescales as in the S2 case, and building a WF model for S1 using non-saturated channels requires alignment of signals in all channels better than 0.01\,$\rm{\mu}$s. This is not achieved in XENON1T as the arrival time of each photon, the PMT time responses, and the length of readout cables are all different. For these reasons, the saturation correction described above is not applied to the S1.

\subsection{Identification of primary and secondary signals}
\label{subsec:peak_indentification}
Secondary signals are defined as signals not directly caused by particle interactions in the LXe. They are associated with light and electron emission induced by S1s or S2s. Depending on the location of the emission we subdivide them into two main types. Gas present in PMTs can be ionized by accelerated electrons between the photocathode and the first dynode~\cite{Barrow:2016doe}, producing after-pulse (AP) signals. Both photo-detachment of electronegative impurities and the photoelectric effect at the metal surfaces of the gate electrode produce electrons within the LXe, that in turn produce spurious S2 signals that we call photoionization (PI) signals~\cite{Aprile:2013blg}.

\begin{figure*}[t]
    \centering
    \includegraphics[width=0.9\textwidth]{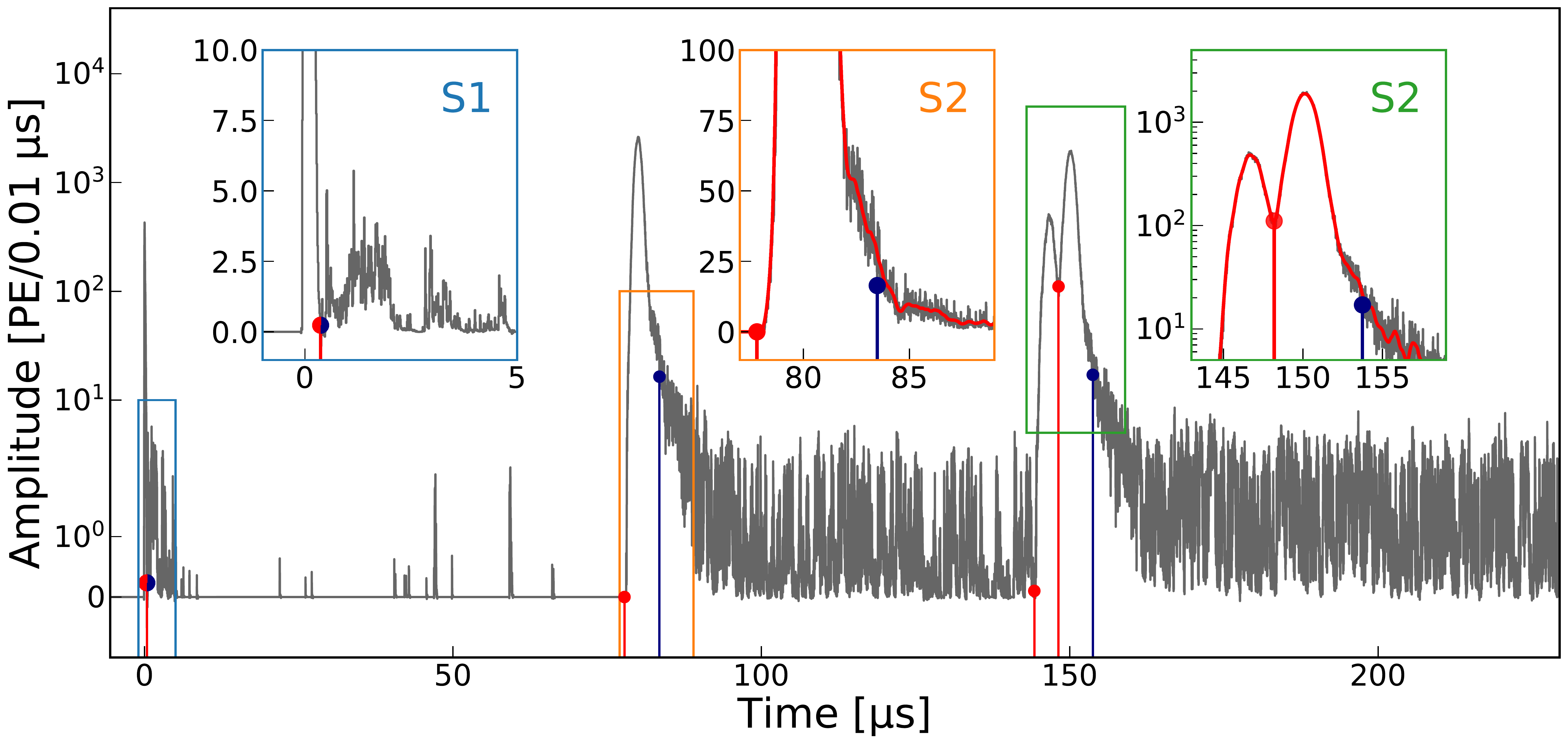}
    \caption{A sample WF (after correction) of a high-energy event induced by a series of Compton scatters in the LXe. The summed WFs are shown as grey lines while the smoothed summed WFs are shown as the overlaid red lines in the insets.  The WF of such an event typically has a narrow S1 peak and a few S2 peaks, each of which is followed by secondary signals from AP and PI processes. The effect of the algorithms on each peak is highlighted by the insets, with the final peak edges shown by vertical lines. The red points represent the local minima that define the end of the S1 signal and the start of each S2 signal. The blue points represent the threshold of 0.13\% of the peak size. While secondary signals are clearly separated from the S1 peak, they overlap with S2 peaks. }
    \label{fig:clustering}
\end{figure*}

Since both AP and PI signals start to appear shortly ($\simeq1$\,$\mu$s) after the primary S1 or S2, they have significant effects on finding the peak boundaries. This leads to sizeable non-linearity and fluctuations in the reconstructed energy. Fig.\,\ref{fig:clustering} shows the S1-S2 signal from a gamma-ray Compton scattering in the LXe after the saturation correction of Sec.~\ref{saturation_correction} is applied. Each S1 and S2 is succeeded by AP and PI. While one can isolate the S1 from secondary signals based on the waveform, the S2s are too wide to separate such secondary signals out. Two algorithms are designed to discriminate and reject those secondary signals, as well as to identify individual interaction sites, using a WF summed over all channels. The two algorithms are complementary to each other. When they suggest splitting at two nearby points instead of the same point, the one closer to the primary signal is chosen as the final peak boundary.

\begin{enumerate}
    \item To minimize the impact of noise, the summed WF (grey lines in Fig.\,\ref{fig:clustering}) is smoothed (red lines) using a locally weighted smoothing method as in \cite{doi:10.1080/01621459.1979.10481038}. Local minima found in the smoothed summed WF are used to define peak boundaries marked as red points in Fig.\,\ref{fig:clustering}. One of them is found in the gap between the S1 and secondary signals defining the end of the S1; two are found at the beginning of the S2 signals to split them from preceding secondary signals; the last one is found between overlapping S2 signals from two interaction sites.
    \item A cutoff on the amplitude is set for each peak to define the extent of its falling edge. The cutoff threshold is placed at the value of a Gaussian function 3-$\sigma$ away from its center, with the height of the Gaussian matching the height of the peak. When the falling edge of the peak falls below this threshold, the peak is truncated in order to detach the tails from AP and PI. Thus, only 0.13\% of the peak area is removed if the peak is Gaussian, as expected from the longitudinal diffusion of the electron cloud~\cite{Sorensen_2011}. Marked as blue points in Fig.\,\ref{fig:clustering}, the cutoff of the S1 is found to coincide with a local minimum; the cutoff points of the S2s split away most of the secondary signals, and their integrated area before the cutoff is approximately proportional to the size of S2.
\end{enumerate}

\subsection{Position reconstruction}
\label{pos_rec}
\begin{figure}[h]
    \centering
    \includegraphics[width=0.46\textwidth]{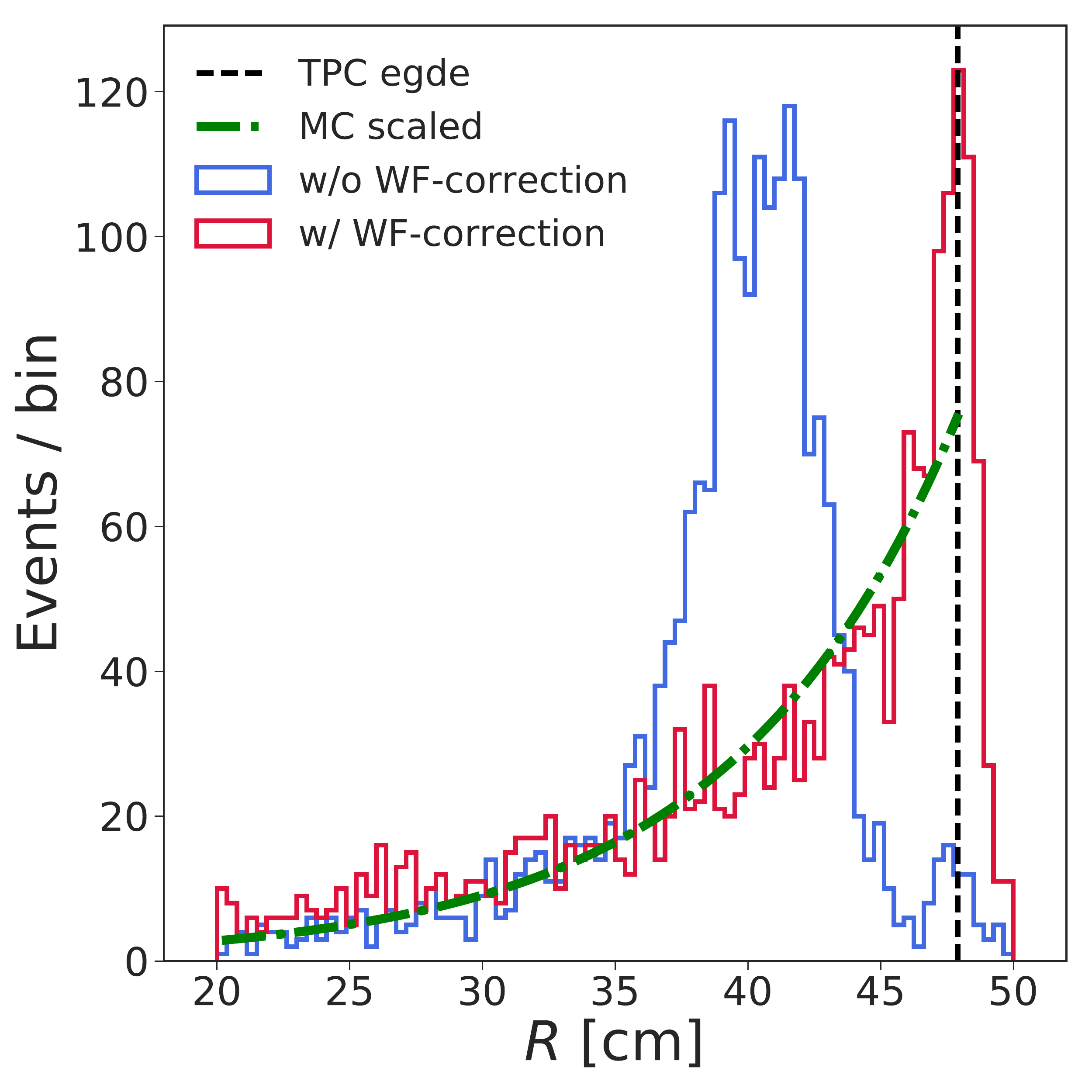}
    \caption{Comparison of the radial position distribution of $^{208}$Tl events from external $^{228}$Th calibration, between data processed with (red) and without (blue) WF correction. Also shown here are the maximum radius of the TPC (black) and the distribution of simulated $^{208}$Tl events (green), scaled to fit the red distribution.}
    \label{fig:th228_calibration_r_distribution}
\end{figure}

The ability to reconstruct the three-dimensional position of events is a key advantage of dual-phase TPCs.
The horizontal coordinates, $x$-$y$, are reconstructed from the S2 light pattern in the top PMT array. Thus, to obtain an unbiased position, the WF correction is applied to the S2 signal. 
Calibration data from an external $^{228}$Th source are used to check the improvement of the position reconstruction induced by the saturation correction. The calibration source is placed at the side of the detector, close to the top of the TPC, which increases the number of saturated events and avoids the field distortion effect as in ~\cite{Aprile:2019bbb}. The radial position distribution of events from the $^{208}$Tl line at 2614.5\,keV, mainly at the edge of the detector, is shown in Fig.~\ref{fig:th228_calibration_r_distribution}. The same position reconstruction method is used, with (red) and without (blue) the saturation correction applied. The distribution of saturation-corrected reconstructed positions shows good agreement with the 48\,cm maximum radius determined by the inner surface of the PTFE reflector, while the distribution without correction shows a significant inward bias.

Similar to the method detailed in~\cite{Aprile:2019bbb}, a feed-forward neural network is used to reconstruct $x$-$y$ coordinates. To improve the precision of the position reconstruction, a deeper network with four hidden layers is constructed using the Keras\,\cite{chollet2015keras} package with the TensorFlow\,\cite{tensorflow2015-whitepaper} backend. The dropout\,\cite{10.5555/2627435.2670313} technique is applied to avoid over-fitting the network to the training set. Compared to~\cite{Aprile:2019bbb}, this neural network improves the position reconstruction precision by $\simeq30$\% and leads to a more uniform response across the detector. Additionally, distortions in the position distribution due to an imperfect drift field are taken into account using the approach presented in~\cite{Aprile:2019bbb}.

\section{Electronic recoil energy reconstruction}
\label{sec:xenon1t_results}
The energy resolution, which is particularly important for the \vzbb-decay sensitivity, can be improved by applying the reconstruction techniques described in the previous sections.
In this section, the calculation of the energy resolution using background data is described for single-site (SS) and for multi-site (MS) interactions.

\subsection{Single and multi-site interactions}
\label{subsec:single_multiple_scatter}
The number of interaction sites of an event is a key feature for discriminating background in the search for rare events. SS interactions encompass potential signals from rare physics processes like dark matter, \vzbb and \vtbb decays. In the latter two cases, the two betas are emitted at the same interaction point. Their penetration length in LXe is less than 3\,mm\,\cite{estar_nist}, indistinguishable with the spatial resolutions of the XENON1T detector. The expected signature is then a single scatter made of two coinciding, unresolved betas. Background contributions for these searches originate from interactions due to beta decays and gamma-rays. MS interactions, mainly due to multiple Compton scatters of gamma-rays (or the coincidence of two gamma-rays happening at the same time), are used to identify and constrain the background components.

\subsection{Combined energy from S1 and S2}
\label{subsec:energy_reconstruction}

A linear, electric field independent relationship between energy and total number of produced quanta (either scintillation photons or ionization electrons) has been established in LXe dual-phase TPCs built for dark matter searches, such as XENON100\,\cite{Aprile_2012}, LUX\,\cite{Akerib:2016qlr}, PandaX-II\,\cite{Ni:2019kms}, as well as LXe TPCs built for \vzbb, such as EXO-200\,\cite{Anton:2019wmi}. The energy transferred in an interaction can be expressed as
\begin{linenomath}
\begin{equation}
\label{eq:energy}
E = (n_{\rm{ph}} + n_{\rm{e}})\cdot W = \left( \frac{\rm{S1}}{g_1} + \frac{\rm{S2}}{g_2}\right) \cdot W \text{ ,}
\end{equation}
\end{linenomath} where $W$ = (13.7$\pm$0.2)\;eV/quantum \cite{Dahl:2009nta} is the average energy needed to produce either scintillation or ionization, and $n_{\rm{ph}}$ and $n_{\rm{e}}$ are the number of emitted photons and electrons. The scintillation photons and ionization electrons are then detected as S1 and S2 signals, with a photon detection efficiency of $g_1$ and charge amplification factor of $g_2$. These are detector-dependent parameters that are determined using mono-energetic peaks, including $^{83\textrm{m}}$Kr, $^{129\textrm{m}}$Xe, $^{131\textrm{m}}$Xe, $^{60}$Co and $^{208}$Tl. We rewrite Eq.\,(\ref{eq:energy}) as
\begin{linenomath}
\begin{equation}
\label{eq:e_lin}
\textrm{QY} = -\frac{g_2}{g_1}\textrm{LY} + \frac{g_2}{W} \text{ ,}
\end{equation}
\end{linenomath} where QY = $\rm{S2}/E$ and LY = $\rm{S1}/E$ are the mean charge yields and the mean light yields at each energy.

\begin{figure}[tp]
    \centering
    \includegraphics[width=0.46\textwidth]{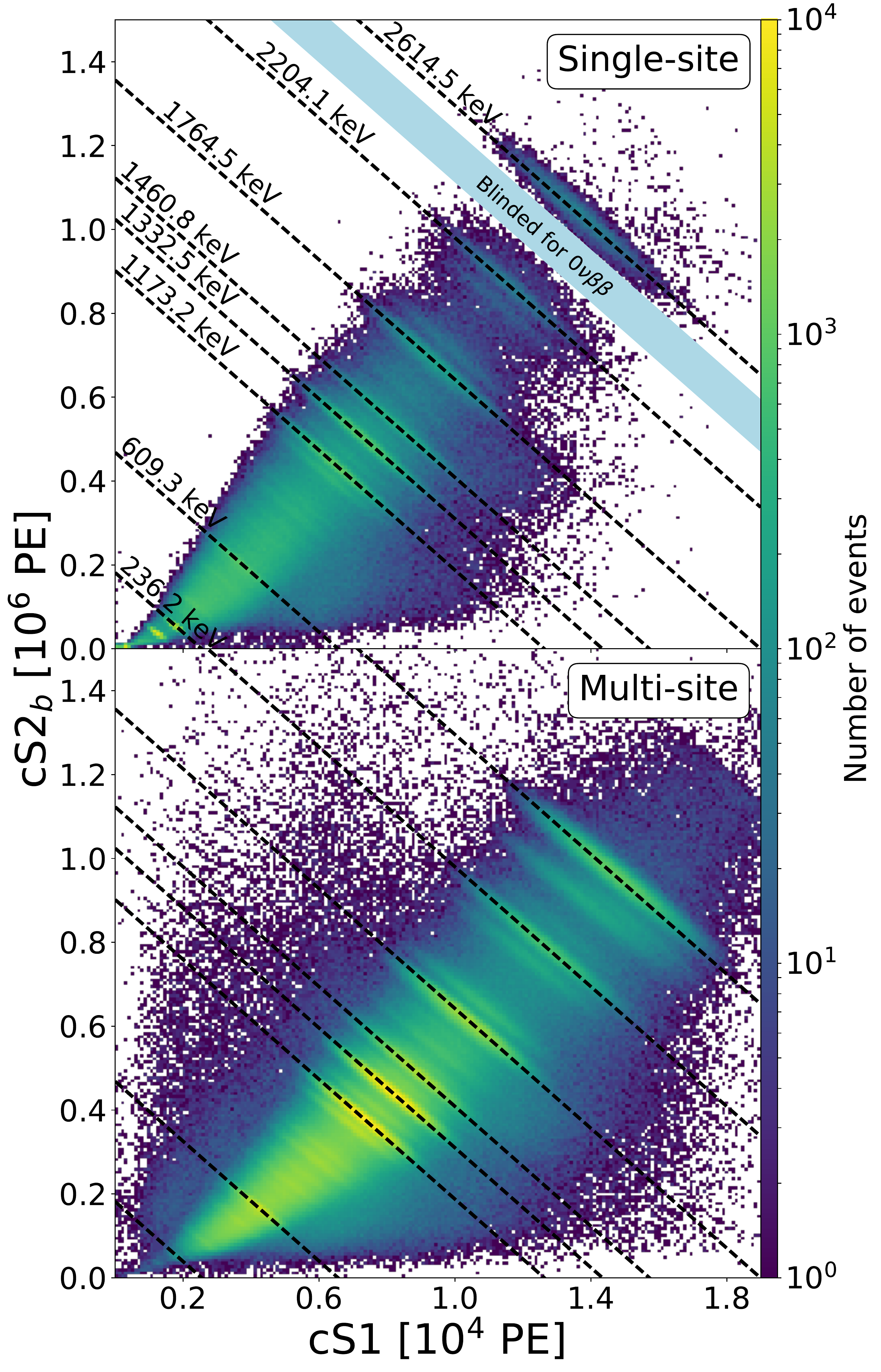}
    \caption{Single-site (top) and multi-site (bottom) background event distributions in corrected S1 (cS1), and corrected S2 bottom (cS2$_b$), space. Mono-energetic photo-absorption peaks of gamma-rays are labelled with their energies and corresponding sources. SS events with energies around $\rm{Q}_{\beta\beta}$ are blinded. }
    \label{fig:2d_bkg_ss}
\end{figure}

Fig.~\ref{fig:2d_bkg_ss} shows the distributions of background events for SS (top) and MS (bottom) interactions. The top PMT array is excluded from the summed S2 size to avoid detection efficiency changing suddenly in the $x$-$y$ plane under the non-operational PMTs. PMTs on the bottom array with large AP rate are also excluded. Leaving those PMTs out doesn't increase the associated statistical fluctuations thanks to the amplification in gaseous xenon. S1 and S2 signals are then corrected with the relative detection efficiencies at different positions derived from the $^{83\textrm{m}}$Kr calibration. The derivation is updated from the approach detailed in~\cite{Aprile:2019bbb}, considering the electric field effect on the $^{83\textrm{m}}$Kr events. In particular, a linear correction depending on the depth of the interaction had to be added for both SS and MS events. This is slightly higher for the MS events due to a larger contribution of the AP and PI to the S2 signals. For a MS event, the combined S1 is corrected with the average of the relative light detection efficiencies at each of the S2s' positions, and weighted by the size of the S2s.

\begin{figure}[htp]
    \centering
    \includegraphics[width=0.46\textwidth]{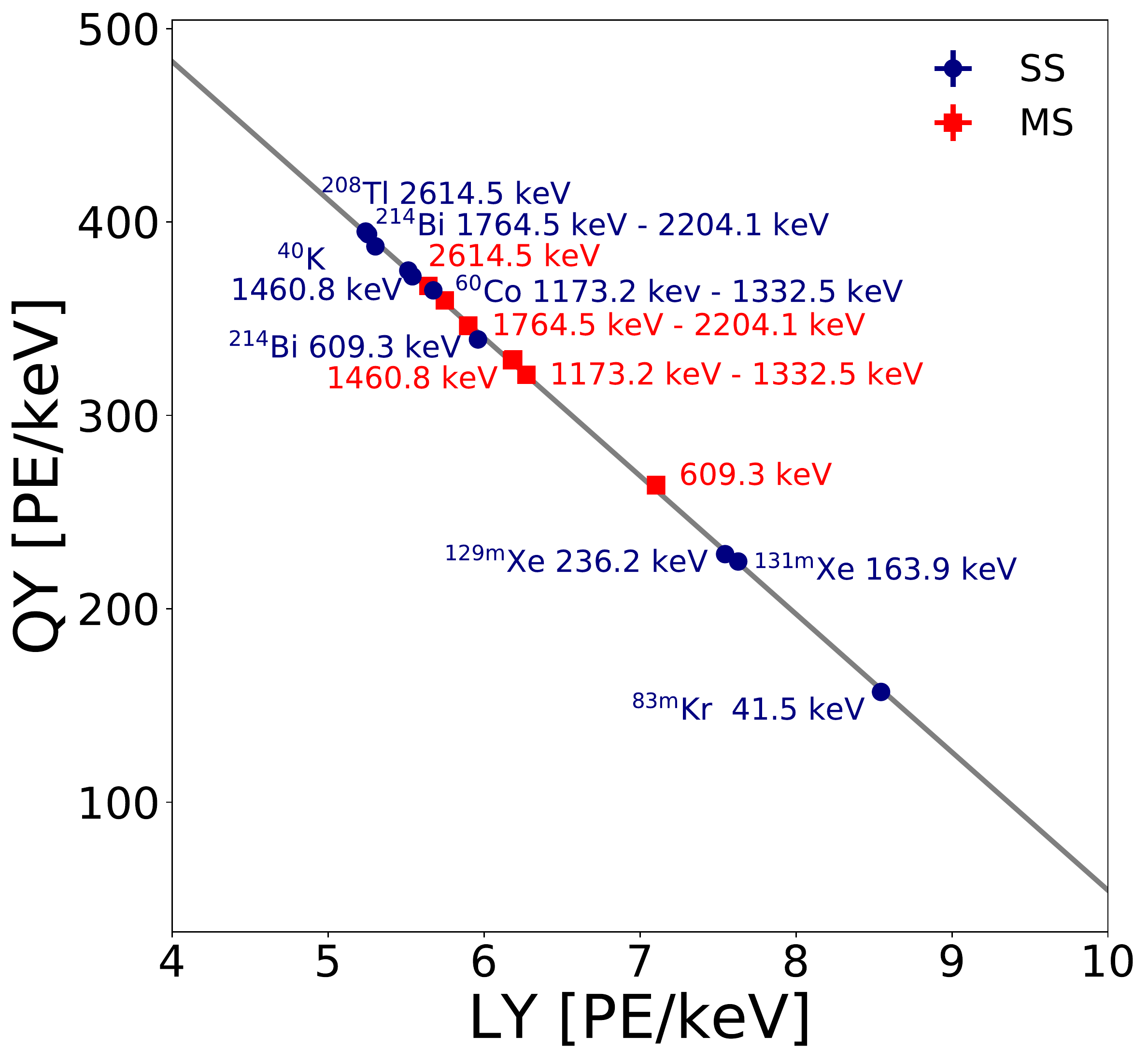}
    \caption{Anti-correlation between the measured light yield (LY) and charge yield (QY) using mono-energetic lines. These data points are from SS (blue) and MS (red) events in the inner 1\,t fiducial volume.  The values are different for SS and MS events at a given energy due to the energy-dependent ion-electron recombination processes. The best linear fit for combined SS and MS data points is overlaid as a solid line (grey).}
    \label{fig:g1_g2_ss}
\end{figure}

\begin{figure*}[htp]
    \centering
    \includegraphics[width=0.9\textwidth]{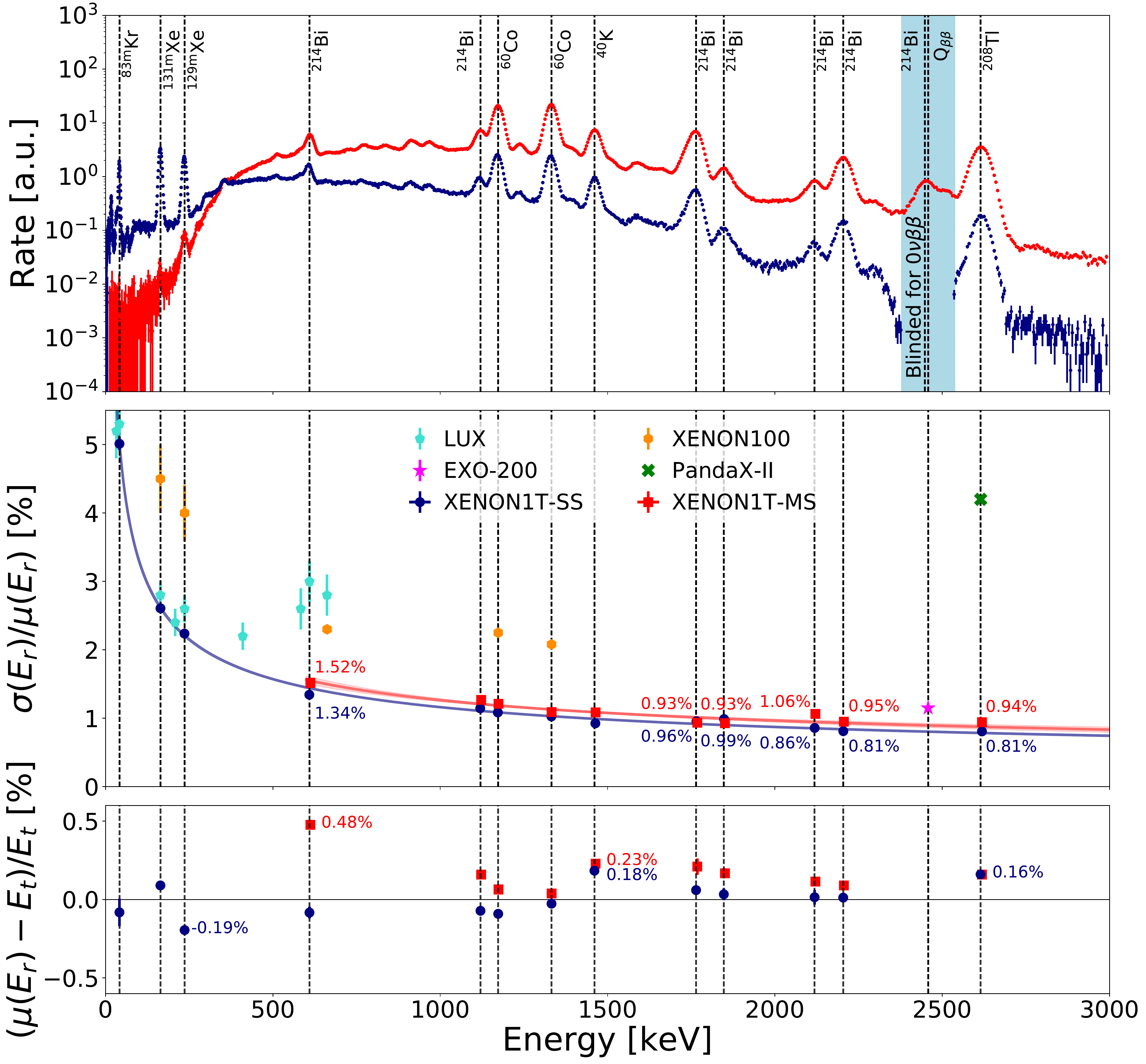}
    \caption{Top: Electronic recoil energy spectra of single-site (blue) and multi-site (red) events in the central 1\ton\xspace fiducial volume of XENON1T. SS events with energies around $\rm{Q}_{\beta\beta}$ are blinded for the search for \vzbb decay. The corresponding decaying isotope for the most visible peaks is labelled with a dashed vertical line. The MS spectrum has a lower rate at low energies due to the fiducial volume selection. Middle: The measured energy resolution for SS and MS events. The SS and MS resolutions as a function of energy are fit with $a/\sqrt{E}+b$ and shown by the blue and red lines, respectively, while the shaded regions cover 1-$\sigma$ statistical uncertainty of the fits. The extrapolated values for the SS are $a$ = (31.3$\pm$0.7) and $b$ = (0.17$\pm$0.02). The resolution of XENON100\,\cite{Aprile_2012}, LUX\,\cite{Akerib:2016qlr}, PandaX-II\,\cite{Ni:2019kms} and EXO-200\,\cite{Anton:2019wmi} are also reported. Bottom: The relative energy shift from the true values for SS and MS events.}
    \label{fig:spectrum_ss}
\end{figure*}

The relative LY and QY are estimated by 2-dimensional Gaussian fits to each monoenergetic peak above the background\,\cite{Aprile_2012}\,\cite{PhysRevB.76.014115}. Fig.\,\ref{fig:g1_g2_ss} shows the relation between LY and QY. At given interaction energies, these measured values are different for SS and MS events due to the energy-dependent ion-electron recombination processes. $g_2$ and $g_1$ depend on the specific characteristics of the detector and on the type of interaction. As SS and MS light and charge yields from ER interactions provide consistent values, they are fitted together. The extrapolated parameters are
\begin{linenomath}
\begin{align}
    \label{eq_g1}
    g_1 &= (0.147 \pm 0.001) \text{\,PE/photon} \text{ ,}\\
    \label{eq_g2}
    g_2 &= (10.53 \pm 0.04) \text{\,PE/electron} \text{ ,} 
\end{align}
\end{linenomath}
The reconstructed energy is then calculated using these values for both SS and MS events.

\subsection{Linearity and resolution of the reconstructed energy}

\begin{figure}[]
    \centering
    \includegraphics[width=0.46\textwidth]{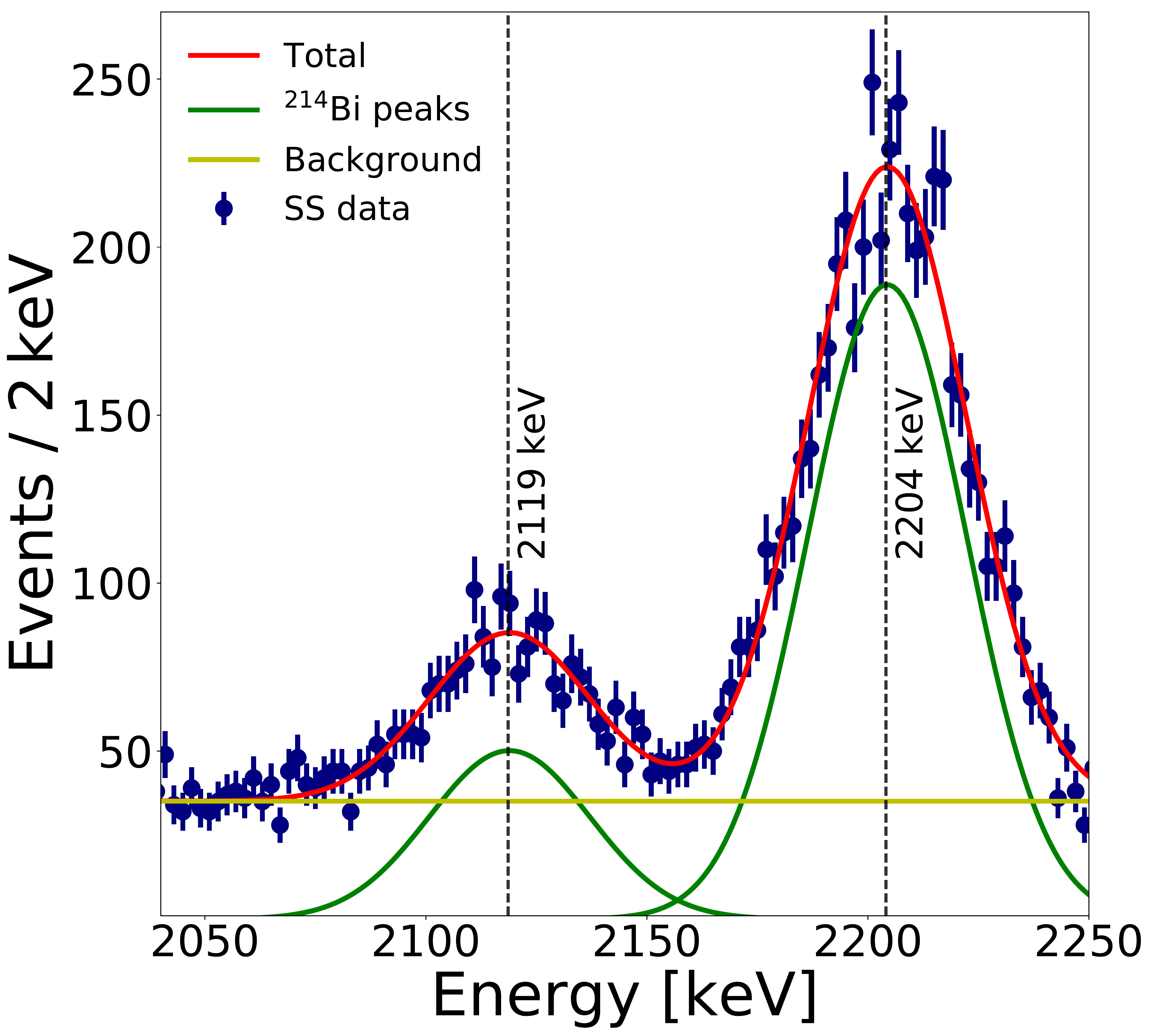}
    \caption{Fit on the $^{214}$Bi peaks above 2\,MeV, $\chi^2$/d.o.f = 113.7/118. The extracted value and standard deviation for the peak at higher energy are $\mu$ = 2204.4\,keV and $\sigma$ = 17.9\,keV, respectively. The computed resolution is then $\sigma/\mu$ = 0.81\%.}
    \label{fig:fit_peak}
\end{figure}

The reconstructed energy spectra for both SS and MS data are shown in the top panel of Fig.\,\ref{fig:spectrum_ss}. Mono-energetic gamma lines from radioactive decays are fitted with Gaussian distributions above a background characterized by a constant or linear function around the peaks. An example is shown in Fig.\,\ref{fig:fit_peak}. In other cases, when the background around the peak is rapidly changing, an exponential function is added to the fit as well.
The fits yield the resolution of the reconstructed energy, $\sigma(E_{\rm{r}})/\mu(E_{\rm{r}})$, and its shift from the nominal value, $(\mu(E_{\rm{r}})-E_{\rm{t}})/E_{\rm{t}}$, the reconstructed energy being $E_{\rm{r}}$ when the true value is $E_{\rm{t}}$, with a mean value of $\mu(E_{\rm{r}})$ and a standard deviation of $\sigma(E_{\rm{r}})$. The shift observed across the entire energy range for both SS and MS data is $\leq 0.4$\%. For comparison, the S2 signals on the bottom PMT array are biased up to -3\% at 2.5\,MeV if the saturation correction is not applied. The excellent linearity of the energy response further ensures that the $g_2$ and $g_1$ calibrated at higher energy are applicable to low energy signals.

The energy resolution of SS data acquired during 246.7 days of dark matter search by XENON1T is (0.80$\pm$0.02)\,\% in one-ton fiducial mass at 2.46\,MeV, to be compared with the 4.2\,\% reported for the dual-phase LXe TPC of the PandaX-II experiment~\cite{Ni:2019kms} and the energy resolution of (1.15$\pm$0.02)\,\% achieved in EXO-200~\cite{Anton:2019wmi}. 
The achieved resolution for MS events at 2.46\,MeV is (0.90 $\pm$ 0.03)\%. The slightly lower resolution from MS data with respect to SS data is due to limitations in the identification, reconstruction and corrections of both the S1 and S2.

\section{Conclusions and outlook}
\label{sec:conclusion}

We have presented signal reconstruction and correction methods designed to improve the energy linearity and resolution at MeV energies in the XENON1T dual-phase TPC. We have devised procedures to correct S2 signals with saturation due to both the digitizers' dynamic range and distortions caused by the non-linear response of the PMT voltage divider circuits and the amplifiers.
We obtained an unprecedented relative energy resolution of 1\,$\sigma/\mu$ = (0.80$\pm$0.02)\,\% at 2.46\,MeV in a drift field of 81\,V/cm. This resolution is mostly limited by fluctuations in the scintillation and ionization signals. The photon detection efficiency $g_1$ determines the fluctuations in the scintillation signal.
The mean electrons' drift length before absorption by electronegative impurities in the liquid determines the fluctuations in the ionization signal. In XENON1T, the mean drift length is $\ge 80\,$cm, leading to a $\simeq30$\% survival probability of an ionization signal at the bottom region of the TPC. This is significantly higher than for the scintillation channel, where the efficiency is $\simeq12$\%. Further improvements in energy resolution can be achieved with larger photosensor coverage and higher quantum efficiency which would reduce the fluctuations in the scintillation signal.

The upcoming XENONnT experiment, an upgrade of XENON1T with a larger TPC and reduced background, is expected to start taking data in 2020. 
Several detector improvements will enhance the energy reconstruction of high-energy events. 
Firstly, the dynamic range of the S2 signal will be extended. The amplifiers of the top PMTs will feature dual gains, a high-gain channel with 10X amplification, and a low-gain channel with a 2X attenuation. 
Secondly, smaller fluctuations in the ionization channel are expected due to a longer mean drift length of electrons before absorption, thanks to a cryogenic LXe purification system with higher circulation speed. Beside the hardware upgrades, the energy reconstruction in XENONnT will still benefit from the WF correction algorithm developed in this work, to address the distortions on the analog signals such as those due to the PMT voltage divider circuits.
The resulting improvement in energy resolution and linearity, coupled with the expected lower background of the new detector, will make it well-suited to search for rare events beyond those expected from dark matter particles, such as the neutrinoless double-beta decay of $^{136}$Xe. 

\begin{acknowledgement}
\label{sec:acknowledgement}
We gratefully acknowledge support from the National Science Foundation, Swiss National Science Foundation, German Ministry for Education and Research, Max Planck Gesellschaft, Deutsche Forschungsgemeinschaft,
Netherlands Organisation for Scientific Research (NWO), Netherlands eScience Center (NLeSC) with the support of the SURF Cooperative, Weizmann Institute of Science, Israeli Centers Of Research Excellence (I-CORE), Pazy- Vatat, Fundacao para a Ciencia e a Tecnologia, Région des Pays de la Loire, Knut and Alice Wallenberg Foundation, Kavli Foundation, and Istituto Nazionale di Fisica Nucleare. This project has received funding or support from the European Union’s Horizon 2020 research and innovation programme under the Marie Sklodowska-Curie Grant Agreements No. 690575 and No. 674896, respectively. Data processing is performed using infrastructures from the Open Science Grid and European Grid Initiative. We are grateful to Laboratori Nazionali del Gran Sasso for hosting and supporting the XENON project.
\end{acknowledgement}

\bibliographystyle{spphys}       
\bibliography{main}   

\end{document}